\documentstyle[11pt]{article}

\newcommand{\eq}{\begin{equation}}
\newcommand{\en}{\end{equation}}

\def\spinst#1#2{{#1\brack#2}}
\def\L{\Lambda}

\newcommand{\NP}[1]{Nucl.\ Phys.\ {\bf #1}}
\newcommand{\PL}[1]{Phys.\ Lett.\ {\bf #1}}

\newcommand{\PR}[1]{Phys.\ Rev.\ {\bf #1}}
\newcommand{\PRL}[1]{Phys.\ Rev.\ Lett.\ {\bf #1}}

\newcommand{\IJMP}[1]{Int.\ J.\ Mod.\ Phys.\ {\bf #1}}

\begin{document}
\begin{titlepage}
\title{
{}~~~~~~~~~~~~~~~~~~~~~~~~~~~~~~~~~~~~~~~~~~~~~~{\normalsize DFTT 23/92}
\\{~}\\{~}\\
The Effective String of ${3D}$ ${Z_2}$ \\
Gauge Theory as a ${c=1}$ Compactified CFT\thanks
{\it Work supported in part by Ministero dell'Universit\`a e della
Ricerca Scientifica e Tecnologica}\\}
\author{ M. Caselle$^a$\thanks{\it email address Decnet=
(39163::CASELLE) Bitnet=(CASELLE@TORINO.INFN.IT)}\ , R. Fiore$^b$ ,
F. Gliozzi$^a$  and S. Vinti$^a$ \\{~}\\
{\footnotesize {\em
$^a $ Dipartimento di Fisica Teorica dell'Universit\`a di Torino}}\\
{\footnotesize {\em
Istituto Nazionale di Fisica Nucleare,Sezione di Torino}}\\
{\footnotesize {\em
Via P.Giuria 1, I-10125 Turin,Italy}}\\
{\footnotesize {\em
$^b $Dipartimento di Fisica, Universit\`a della Calabria}}\\
{\footnotesize {\em
Istituto Nazionale di Fisica Nucleare, Gruppo collegato di
Cosenza}}\\
{\footnotesize {\em
Arcavacata di Rende, I-87036 Cosenza, Italy}}\\}
\date{ }
\maketitle
\begin{abstract}
We report on a high precision Montecarlo test of the
three dimensional Ising gauge
model at finite temperature. The string tension $\sigma$ is
extracted from
the expectation values of correlations of Polyakov lines. Agreement with
the string tension extracted from Wilson loops is
found only if the quantum fluctuations of the flux tube are properly
taken into account. The central charge of the underlying conformal field
theory is $c=1$.
\end{abstract}
\end{titlepage}

\vfill
\eject

\setcounter{page}{1}

{\bf 1. Introduction}
\vskip .6 cm
In the last years
lot of efforts have been devoted to extract interquark potential from
lattice gauge theories (LGT's) looking at the expectation values of Wilson
loops or (correlations of) Po\-lya\-kov lines in Montecarlo
simulations. Besides the important issue of obtaining reliable
values of physical observables like the string tension,
these simulations also allow to study the physical
nature of the potential. In particular, more and more convincing
evidences have been reported in last years of the so called ``string
picture'' of the interquark potential: quark and antiquark linked
together by a thin fluctuating  flux tube~\cite{conj}.
\vskip.3cm

Roughly speaking, there are two independent ways to check this picture.

The first one is to try to observe directly the flux tube, for instance
as a stable semiclassical configuration, surviving
cooling~\cite{cooling} or looking at the space distribution of colour
flux~\cite{wosiek}.
\vskip .3cm

The second one, followed in this paper, is to look at the finite size
effects due to quantum string fluctuations, which, in finite geometries,
give significative, measurable contributions to the interquark potential.
This second approach traces back to the work of L\"uscher, Symanzik and
Weisz~\cite{luscher} and has interesting connections with the conformal
field theory (CFT) approach of two-dimensional models developed in the
last years\cite{cft}.
\vskip.3 cm

This approach has particularly relevant consequences if one studies
Po\-lya\_kov lines in LGT's at finite temperature, because in this case
the finite size corrections (which we shall call from now on ``string
contributions'') have a very specific behaviour, due to the choice of
boundary conditions.
\vskip.3 cm

The existence of the string contributions and their explicit
asymptotic form can
be deduced from a purely theoretical point of view as consistency
conditions of non-perturbative descriptions of the infrared behaviour of
gauge theories.
\vskip.3cm

These corrections seem necessary to fit the data of
Montecarlo simulations \cite{n89,n90}. In order to give further evidence
of this fact, we shall compare our theoretical predictions with a new
set of high precision data on the Ising gauge model at finite (but low)
temperature.
In particular, taking into account the non trivial topology of the
finite temperature model,  we shall be able to estimate the conformal
anomaly $c$ of the CFT which describes the long distance behaviour of the
transverse displacements of the string : it turns out to agree with the
value $c=1$ suggested by the naive bosonic string picture.
\vskip.3 cm

This paper is organized as follows: after a general introduction
to the finite temperature gauge models (sect.2), in sect.3 we motivate our
approach and show the general features of contributions coming from the
quantum fluctuations of the flux tube.
Section 4 is  then devoted to the explicit description of the finite
temperature string corrections .
In sect.5 we describe our simulation and discuss the results.

\vskip 0.3cm

{\bf 2. Finite temperature gauge theories: general setting and notations}
\vskip .3 cm
\vskip .3 cm
The partition function of a gauge theory with gauge group $G$
regularized on a lattice is
\eq
Z=\int\prod dU_l(\vec x,t) \exp\{-\beta\sum_p\, Re\, Tr(1-U_p)\}~~~,
\en
where $U_l(\vec x,t)\in G$ is the link variable of position $(\vec x,t)$
and direction $l\in\{x_1,..x_d,t\}$
and $U_p$ is the product of the
links around the plaquette $p$.

Let us call $N_t$ ($N_s$) the lattice size in the
time (space) direction (we assume for simplicity $N_s$ to be
the same for all the space directions). A
$(N_s)^dN_t$ lattice can then be interpreted as representing
a system with a finite volume $V=(N_sa)^d$ and a finite temperature
$T=1/L=1/N_ta$ where $a$ is the lattice spacing.
Lattice simulations with non-zero temperature
are obtained imposing periodic boundary conditions in the time direction.
\vskip.3 cm

The order parameter of the finite temperature deconfinement transition
 is the Polyakov line, $i.e.$  the trace of the ordered product of all
 time links with the same space coordinates and it is
 closed owing  to the periodic boundary conditions in the time direction:
\eq
P(\vec x)=Tr\prod_{z=1}^{N_t}U_t(\vec x,z)~~~.
\en
The vacuum expectation value of the Polyakov line is zero in the
confining phase and acquires a non-zero expectation value in the
deconfined phase. The value $\beta_c(T)$ of this deconfinement
transition is a function of the temperature, and defines a new
physical observable $T_c$.
\vskip.3 cm

The interquark potential can be extracted by looking at the correlations
of Polyakov lines in the confined phase.
The correlation of two lines $P(x)$  at
a distance $R$ and at a temperature $T=1/L=1/N_ta$ is given by

\eq
\langle P(x)P^\dagger(x+R) \rangle = {\rm e}^{-F(R,L)}~~~,
\label{polya}
\en
\noindent
where  the free energy $F(R,L)$ is expected to be described, as a
first approximation,  by the so called ``area law'':

\eq
F(R,L)\sim F(R,L)_{cl}=\sigma L R + c(L)
{}~~~,\label{area}
\en
where $c(L)$ is a  constant depending only on $L$.

The observable (\ref{polya}) is similar to the expectation value of an ordinary
 Wilson loop except for the boundary conditions, which are  in this case
fixed in the space directions and periodic in the time direction. The
 resulting geometry is that of a cylinder, which is topologically
different from the rectangular geometry of the Wilson loop.

\vskip 0.4cm

{\bf 3. The role of the quantum fluctuations of the flux tube}
\vskip .3 cm
\vskip .3 cm

According to an old, long standing
conjecture~\cite{conj},
the area term is only the dominant (classical) contribution of
the effective string which should describe the infrared behaviour of the
gauge theory (this explains the subscript ``$cl$'' in eq.(\ref{area})).
The effective, non critical strings
are (as it is well known) intractable in more than one dimension.
Mainly as a consequence of that, it has become a common
habit in LGT's to ignore the quantum corrections implied by the effective
string hypothesis, only retaining the classical (area) contribution in
fitting the data of Montecarlo simulations. We shall see below the
range of validity and the consequences of this ``zero order
approximation''.
\vskip.3 cm

A way out of this problem is to realize that, for large quark
separations, we do not need to describe the full complexity of the
effective string theory, but only the degrees of freedom which
survive at large distance.
\vskip .3 cm

Indeed the notorious troubles of the non critical strings ( the
breaking of the Lorentz invariance or the coupling to the Liouville
modes) asymptotically disappear at large distances \cite{olesen}.
In such a region the transverse degrees of freedom of
the string are described by the free, massless bosonic fields
of  an ordinary two-dimensional CFT.

The corresponding partition function $Z(R,L)$ will be the first
order quantum contribution of the effective string to the Polyakov line
correlation (or Wilson loop) expectation value. Consequently, taking into
account the corresponding free energy $F(R,L)_q=-\log Z(R,L)$,
we modify eq.(\ref{area}) as follows:

\eq
F(R,L)\sim F(R,L)_{cl}+ F(R,L)_{q}=\sigma L R + c(L)
- \log{Z(R,L)}
{}~~~.\label{a+q}
\en

Due to the recent progress in CFT's~\cite{cft}
 we can study the behaviour of
$F(R,L)_q$ as a function of $R$ and $L$
in a general way. Indeed any two dimensional
CFT is completely described once the conformal anomaly $c$,
the operator content $h_i$ and the operator product algebra (or the
fusion algebra which equivalently encodes all the fusing properties
of the CFT) are given; then it is easy to show that $F(R,L)_q$ only
depends on the adimensional ratio $z=2R/L$. It is possible to
give asymptotic expressions for $F(R,L)_q$ in the $z\gg 1$ and
$z\ll 1$ regimes:

$z\gg 1$~\cite{bcna}
\eq
F(R,L)_q\simeq-(d-2)~\tilde c \frac{\pi R}{6 L}~~~,
\label{zbig}
\en
where $d$ is the space time dimensions of the gauge theory and
$\tilde c=c-24h_{min}$ is the effective conformal anomaly \cite{ISZ};
 $h_{min}$ is the lowest conformal weight of the physical states
propagating along the cylinder. In the case of unitary CFT's $h_{min}=0$
(unless special boundary conditions are chosen)
and $\tilde c$ coincides with the conformal anomaly $c$;

$z\ll 1$~\cite{cardy}
\eq
F(R,L)_q\simeq-(d-2)~\hat c \frac{\pi L}{24 R}~~~,
\label{zsmall}
\en
where $\hat c=c-24h_{\alpha,\beta}$ and
$h_{\alpha,\beta}$ is the lowest conformal weight compatible with the
boundary conditions  $\alpha$ and $\beta$ at the two open  ends of the
cylinder.

If the CFT is exactly solvable, namely if the whole operator content is
known, one can explicitly write the free energy for all values of $z$,
which will smoothly interpolate between the two asymptotic behaviours.
\vskip .3cm

In the Wilson loop case, since all the boundaries of the loop are open
(rectangular geometry), we only have the behaviour given by
eq.(\ref{zsmall}), with a partition function $F(R,L)_q$  symmetric
under the exchange $R\leftrightarrow L$.
\vskip .3cm

An important role in this construction is played by the modular
transformations. All the partition functions can be written as
power expansions in $q=\exp({2\pi i \tau})$, with $\tau=iz$
for Polyakov line correlations or $\tau=i\frac{R}{L}$ for Wilson loops.
Modular transformations allow to extend these expansions in the whole
$\tau$ plane. In particular we will be interested, in the following, in
the $\tau \to -1/\tau$ transformation.
\vskip .3cm

In the Wilson case, this
transformation is a symmetry, because it exchanges $R$ and $L$.
We can use this symmetry by choosing
for instance $L\geq R$, and $\tau=iL/R$. Here $L$ plays the role of time
 and  the interquark
potential $V(R)$ we want to extract from the data is defined in the
limit: $V(R)=\lim_{L\to\infty}F(R,L)/L$.
\vskip .3cm

In the Polyakov case the situation is different: $L$ and $R$ have
a  different meaning and the modular transformation
$\tau\to-1/\tau$ allows us to move from the region in which $2R>L$ to
that in which $2R<L$. What is new is that, due to the modular
transformation, in these two regions the string corrections have, as we
have seen above, different functional forms.
While in the region in which $2R<L$ the dominant
contribution is, like in the Wilson case, of the type $1/R$,
in the region
in which $2R>L$ the dominant contribution is proportional to $R$,
and acts as a finite size correction of the string tension. This
behaviour will play a major role in the following.

\vskip 0.3cm

Let us now look at the other side of the problem: the data coming from
computer simulation. Indeed, in the last years, as a consequence of the
improvement in precision and size of the Montecarlo simulations of LGT's,
serious problems arose in the interpretation of data. These problems
can be summarized into three items:

a) scaling deviations,

b) disagreement between data and the ``area, perimeter and
constant'' law,

c)disagreement
between Wilson loop and Polyakov line correlation estimates of the string
tension.

Let us briefly discuss these items.
\begin{description}
\item{a)}
It is well known that the most safe way one has to separate
physically meaningful quantities from lattice artifacts is to test their
correct scaling behaviour near the continuum limit, and this is the only
ultimate check of the whole Montecarlo procedure. Indeed, one of the
earliest success of LGT's was the identification of
such a scaling behaviour in the string tension~\cite{creutz}.

However the improvement in the precision and size of lattice
simulations led to the observation of significative deviations from
asymptotic scaling~\cite{dev} of the string tension extracted from
Wilson loops. In~\cite{n89,n90} we have shown that
taking into account string corrections a good scaling behaviour could be
obtained for the string tension extracted from  Wilson loops both in
$4D$ $SU(2)$ and $SU(3)$ models~\cite{n89} and in $3D$ $Z_2$and $Z_5$
models~\cite{n90}.

We shall show in this paper a similar scaling
behaviour for the string tension $\sigma$ extracted from Polyakov line
correlations  at finite temperature. What is peculiar of the Polyakov
case is that the string corrections for $z>1$  act simply  as a
renormalization of  $\sigma$ (see eq.(\ref{zbig}));
as a consequence, in this region the scaling behaviour of $\sigma$
can be obtained even if one  neglects string corrections .

Owing to the fact that in the actual simulations the  size $L$ is
always rather small, it is enough to cut off few short range
(i.e. small $R$) data to reach $z>1$ and hence scaling.

This probably explains why in the last years Polyakov line correlations
have been so popular in estimating the string tension and have almost
substituted Wilson loops.

\item{b)}
The  seeming absence of scaling of the string tension extracted
from the Wilson loops or Polyakov line correlations at short distance is
mainly due to a  sizable violation of
the expected area law  of the Montecarlo data.
 As a matter of fact, if one
tries to fit also the short range data with the area law,
the resulting $\chi
^2$ is unacceptably high. Moreover, setting a lower cut-off on the
fitted data, a definite systematic trend can be seen in the resulting
string tension, which decreases as the the lower bound increases (see
fig.1).
Let us stress that this behaviour is impressively universal: it can be
seen in the string tension extracted both from Wilson loops or Polyakov
line correlations, either in non abelian $SU(2)$ and $SU(3)$ gauge
theories in four dimensions~\cite{n89}, or in the $Z_2$ model in three
dimensions~\cite{n90}, . Most probably, this systematic trend is the
main reason of the fact that, as simulations were performed on lattices
of larger and larger size, and bigger Wilson
loops could be reached, the published values of, say, the pure QCD
string tension became smaller: from the $\frac{\sqrt{\sigma}}{\Lambda}
\sim 140$ of ref.~\cite{creutz} in 1980 to the
$\frac{\sqrt{\sigma}}{\Lambda}\sim 80$  estimate of
ref.s~\cite{ding,n89,bali,michael}.

If one goes on in eliminating short range data,
acceptable confidence levels are eventually reached when all the data
below a suitable threshold $R_\sigma$ are ignored.

Unfortunately this
agreement with the area law is only apparent; indeed  if we
 increase $R_\sigma$  further, we still observe a decreasing string
tension, a trend which is hidden by the error bars (linearly
growing with the distance) thus allowing apparently good $\chi^2$ (see
fig.1). However the values of $\sigma$ extracted at these values of
$R_\sigma$ do not scale as a universal function of $\beta$ as they
should do in the asymptotic scaling region.

Notice  that this behaviour of $\sigma$ as a function of $R_\sigma$
is not a lattice artifact, since it scales, as we shall see below.
In particular, we shall account for this behaviour  by considering the
contributions of the quantum fluctuations of the string.

Let us conclude this item with a  remark on the physical scales involved
in these quantum fluctuations.
In principle, one could ignore all these complications
simply by looking at very large Wilson loops. This corresponds, in the
continuum limit, to study quark-antiquark pairs with large separations.
Taking as an example SU(3) in $4D$ and physical unities, we can ignore
string corrections for distances greater than $\sim 1.$ fm., but they
are definitely relevant at distances ($\sim 0.2$ fm.) typical of heavy
quarkonia ($c\bar c$, $b\bar b$).

\item{c)}
In this paper we study the Ising gauge model at finite, but low,
temperature, where all the effects of the deconfinement transition can
be neglected\footnote{These effects of extreme interest for the
effective string picture, since the quantum string fluctuations are
more relevant near the deconfinement point. However one should take
into account higher order contributions. We have chosen to avoid these
complications in order to make more definite  the comparison
between Montecarlo data and our predictions.}.
In this region the string tension measured with Wilson
loops should be the same of the zero temperature case. In fact,
extracting $\sigma$ from
Wilson loops orthogonal to the time direction, we found that its
value, (compared with the zero temperature results of
ref.~\cite{n90}\footnote{Actually, owing to the finite size $(32^3)$ of the
lattice, also the data of ref.~\cite{n90} should be considered,
roughly speaking, as
non-zero temperature estimates of the string tension.
To be precise in~\cite{n90}  the range of
temperatures was  from 0.35$\L$ at $\beta=0.740$ to 0.61$\L$
at $\beta=0.7525$, where the scale $\L$ is defined in eq.(\ref{tc}).})
is unaffected, within statistical errors, by the
compactification of the time direction. On the contrary, if we extract
the string tension from Polyakov line correlations, a different value is
obtained, unless string corrections are taken into account. Indeed this
is a well known phenomenon: the string tension extracted from Polyakov
line correlations is systematically lower than that extracted from
Wilson loops (see for instance~\cite{gao}). The
difference between the two observables is only in the boundary
conditions: these do not affect the area term (hence we expect the same
value of $\sigma$), but they change the string corrections.
As we noticed above, in the Polyakov case,
 for large values of the interquark
distance we must subtract to  $\sigma$, the term
$\frac{\tilde c(d-2)\pi}{6L^2}$, which  explains the disagreement
between the measured values~\cite{gao}.
Our analysis on the $3D$ Ising gauge model agrees with a value
$\tilde c=1$ for the conformal anomaly.

\end{description}

The simplest possible choice for the effective string is given by the
Nambu-Goto action, which is obtained by identifying the flux tube with
the string. The quantum fluctuations of this string can be described,
for large quark separations, by a gaussian model with $d-2$ free bosonic
fields that we shall call in the following the free bosonic string.
This is the simplest possible $2D$ CFT and can be solved exactly.

\vskip .3 cm
The functional integration over the $(d-2)$ bosonic fields
can be done using the $\zeta$
function regularization, and strongly depends on the topology of the
world-sheet~\cite{minami,flensburg,fsst,nordita}.
One finds $\tilde c=\hat c=c=d-2$ for the effective conformal anomalies
defined in eq.s(\ref{zbig},\ref{zsmall}).
 Taking into account this
contribution, we get a definite improvement in the description of the
data and  the above mentioned discrepancies partially disappear.
In particular the problem of point c) is completely solved
 as we shall show in the following.

Notice  that this
improvement is quite general and works for gauge theories both in three
and four dimensions, both with discrete and with continuous groups.
\vskip .3 cm

The free bosonic string can be considered as a sort of mean field
approximation of the proper effective string description and as such,
for instance,
predicts the same (mean field) critical index $\nu=1/2$ near the
deconfinement transition for all gauge theories (see
ref.~\cite{damgaard} for a discussion of this point).

There are three features of the
bosonic string which explain in which sense it is an approximation of
the physical flux tube:
\begin{description}
\item{i)} The bosonic string can freely self-overlap, a feature which
does not agree with our intuition on gauge theory coming both from strong
coupling expansions ( in a $Z_2$ gauge theory, each plaquette must
appear at most once in any
strong coupling diagram) and from semi-classical descriptions of the flux
tube.

\item{ii)} The bosonic string has zero thickness, while the flux tube
has a nonzero, measurable thickness~\cite{wosiek}.

\item{iii)} The bosonic string predicts, for the short distance behaviour
of  $\sigma$ ~\cite{alvarez,arvis}:
\eq
\sigma(R)=\sigma(\infty)\sqrt{1-\Bigl({R_c \over R}\Bigr)^2}~~~.
\label{root}
\en
The term that we can obtain using CFT's (see sect.4) is only the first
order one in the $1/R$ expansion of eq.(\ref{root}). Consequently we expect
that it underestimates the true contribution and that, once it has been
subtracted, the decreasing behaviour of $\sigma(R)$ should be weaker,
but still present.
This is exactly the opposite of what we observe. Both in
the case of Wilson loops (see fig.4 of ref.\cite{n90}) and in the case
of Polyakov line correlations (see fig.1 of the present paper), once bosonic
string corrections are taken into account, $\sigma(R)$ has the wrong
shape.
\end{description}

The free bosonic string is a good description of the
effective string at large distances and low temperatures: it works
for values of $R$ and of the temperature such that the finite thickness
of the flux tube can be ignored, and the self-overlapping
probability is negligible.

\vskip .3 cm
A simple way to take into account the presence of a new scale
(namely the finite thickness of the flux tube) in the problem, is to
compactify into a circle the bosons describing the transverse
displacements . There are several ways to look at this
compactification: one may either think that the string is not any more
identified with the flux tube but wraps around it~\cite{n91a},
or notice that a compactified boson can be fermionized and hence
one can look at the model as a  fermionic string
description~\cite{n89,n90,n91a}
 or, using the connection between critical $O(n)$ models and
compactified bosons, one can discuss the string corrections in term of self-
avoiding walks~\cite{n91c}.
\vskip.3cm

All these approaches capture some of the
physical content of the model, and can give some semiclassical and
intuitive justifications, but let us stress again that the ultimate
reason for introducing a compactification radius is only to set a
new scale in the model. This scale does not represent a new degree of
freedom, since general consistency requirements fix the compactification
radius to be exactly $1/4$ in conventional units ~\cite{n91a}
and select the operator content
of the underlying CFT~\cite{n91a,n91c}.
\vskip.3cm

The compactification on a circle of rational radius
leads to a simpler description of the CFT in terms of Dirac fermions
with a finite number of spin structures ~\cite{n89,n90,n91a}. In the
following we shall call this model ``fermionic string''.
As expected, the large distance behaviour of the fermionic string is the
same of that of the free bosonic string. In particular, one can show
that again $\tilde c=c=1$, hence both strings predict the same gap
between Polyakov and Wilson estimates of $\sigma$. The
difference between the two models becomes sensible when short distance
data (of
the order of the flux tube thickness) are taken into account: in fact
it can be shown that now $\hat c=(d-2)/4$. In this
region the area law corrected with fermionic string contributions
gives definitely better $\chi^2$'s, and reliable scaling values of the
string tension can be extracted (see for instance ref.s \cite{n89,n90}).
Finally,
a further signature of the better behaviour of the fermionic
corrections with respect to the bosonic ones is given by the shape of
the string corrected function $\sigma(R)$ which, as expected, is still a
weakly decreasing function of $R$ (see fig.1 for the Polyakov line
correlations and fig.4 of ref.~\cite{n90} for Wilson loops).

In this paper we shall confirm this picture also in the finite
temperature case.

\vskip 0.4cm
{\bf 4. String corrections in the finite temperature geometry}

The case of the free bosonic string was discussed
in ref.s~\cite{minami,flensburg} and with a different approach
in ref.\cite{fsst} (while the calculation for the rectangular geometry can
be found in~\cite{nordita}). The result is, for $d=3$,

\eq
F_b(R,L)=\log\left({\eta(\tau)}\right)
\hskip0.5cm
;\hskip0.5cm {-i}\tau={L\over 2R}~~~,
\label{bos}
\en
\noindent
where $\eta$ denotes the Dedekind eta function:
\eq
\eta(\tau)=q^{1\over24}\prod_{n=1}^\infty(1-q^n)\hskip0.5cm
;\hskip0.5cmq=e^{2\pi i\tau}~~~,\label{eta}
\en
and $R$ is the distance between the two Polyakov lines.
\vskip.3cm

We list below for completeness the
power expansions in the two asymptotic  regions:

\begin{description}
\item{$2R<L$}
\eq
F_b(R,L)=-\frac{\pi L}{24 R}
+\sum_{n=1}^\infty \log (1-e^{-\pi nL/R})~~~,
\en

\item{$2R>L$}
\eq
F_b(R,L)=-\frac{\pi R}{6 L}+\frac{1}{2} \log\frac{2R}{L}
+\sum_{n=1}^\infty \log (1-e^{-4\pi nR/L})~~~.
\en
\end{description}

Following
the same line of reasoning, one can also obtain the string correction
in the case
in which the transverse degrees of freedom of the string are described
by fermionic fields\footnote{This was done for rectangular geometry
(Wilson loops) in ref.s~\cite{n89,n90} in the case
of Majorana fields (Ising model) and for
Dirac fermions (c=1 conformal field theory) with arbitrary
boundary conditions.}. We shall list here, for completeness, the
results for generic values of the compactification radius $r$, and then
concentrate on the value $r=1/4$.
\vskip.3cm

The string is described by a Dirac spinor $\psi(\xi_1,\xi_2)$
$(0\leq\xi_1\leq R , 0\leq\xi_2\leq L)$ with components
$\psi_\pm(\xi_1,\xi_2)=\psi_\pm(\xi_1\pm\xi_2)$, associated to
the transverse dimension.

The value of the quantum correction depends
on the boundary conditions of $\psi$ on the two fixed edges of the
cylinder (the Polyakov lines). Defining a boundary phase
$\alpha={1 \over N}$ through the relation
\eq
\psi_\pm(\xi_1+2R,\xi_2)=e^{\pm 2\pi i /N}
\psi_\pm(\xi_1,\xi_2)~~~,
\en
\noindent
one has
\eq
F_{f,N}(R,L)=
{-}\log{\sum_{m=0}^{N-1}{\vartheta\spinst{m/N-1/2}
{1/N-1/2}(0|\tau)\over \eta(\tau)}}
;\hskip0.3cm {-i}\tau={2R\over L}~~~,
\label{ferm}
\en
where $\vartheta\spinst{\alpha}{\beta}(0|\tau)$ is the Jacobi Theta
function with characteristic $\spinst{\alpha}{\beta}$.
One can write explicitly $\vartheta\spinst{\alpha}{\beta}(0|\tau)$
in terms of an infinite product in analogy with eq.(\ref{eta}):
\eq
\frac {\vartheta\spinst{\alpha}{\beta}(0|\tau)}{\eta(\tau)}
=q^{{\alpha^2\over2}-{1\over24}}
\prod_{n=1}^\infty
(1+q^{n+\alpha-{1\over2}}e^{2\pi i \beta})
(1+q^{n-\alpha-{1\over2}}e^{-2\pi i \beta})~~~.
\en

Given the above expressions the series expansions for $z<1$ and $z>1$
are, in the case of N=4:

\begin{description}
\item{$2R<L$}
\eq
F_{f,4}(R,L)=-\frac{\pi L}{96 R}
-\log [4(1+q+2q^2+\cdots)]
\label{zs}
\en
with $q=exp(-\frac{\pi L}{R})$~~~,

\item{$2R>L$}
\eq
F_{f,4}(R,L)=-\frac{\pi R}{6 L}
-\log (1+\sqrt{2+\sqrt{2}}q^{1/32}+\sqrt{2}q^{1/8}+\cdots)
\label{zb}
\en
with $q=exp(-\frac{4\pi R}{L})$~~~.

\end{description}

Notice that while the bosonic and the fermionic  strings have
quite different behaviours in the region $2R<L$ due to the different
operator content, they give {\sl exactly the same dominant contribution}
in the region $L<2R$ since they have the same conformal
anomaly $c=1$ and both are unitary CFT's. In this regime the only
relevant difference between the two strings is  a
subdominant contribution of the type $\log R$: this is present
only if the boundary conditions of the transverse degrees of freedom
 are periodic (like in the case of the
bosonic string) and is forbidden if they are non-trivial (like in
the fermionic case).

\vskip .5cm
The above listed contributions are, as a matter of fact, only the first
order terms of the string corrections to the interquark potential. They
are expected to be valid in a window far enough from the roughening
transition so as to allow the string to fluctuate, but not too near to
the deconfinement transition where the string delocalizes, and the
higher order effects due to self-overlapping become important.
\vskip .5cm

{\bf 5. Comparison with Montecarlo data}

In order to measure the conformal anomaly $c$ we made extensive high
statistics simulations of a $3D$ Ising lattice gauge theory,
 extracting Wilson loops and Polyakov line correlations in a
wide range of values of the ratio $R/L$.

We used the standard action
\eq
S=-\beta \sum_{p}U_p~~~,
\en
where $U_p$ denotes the product of $Z_2$ variables associated to the
links belonging to the plaquette $p$. This model is known to have a
roughening transition at $\beta_r=0.4964$ and a zero temperature
deconfinement transition at $\beta_c(N_t=\infty)=0.7614$~\cite{z2beta}
which is of second order. The
corresponding critical index is $\nu\sim 0.63$~\cite{nu}.
The scaling region of the finite temperature $T=1/N_ta$
 deconfinement transition starts approximatively at
$N_t=4$. Data obtained with the Montecarlo renormalization group
approach~\cite{wansleben} for $N_t=4$ and $N_t=8$ give
$\beta_c(N_t=4)=0.7315(5)$ and  $\beta_c(N_t=8)=0.7516(5)$ which
correspond to the following value of the critical temperature $T_c$
\eq
T_c/\L\equiv T_c a(\beta_c(\infty)-\beta_c(N_t))^{-\nu}=2.3\pm0.1 ~~~,
\label{tc}
\en
which is also in agreement with an independent
prediction based on the N=2 supersymmetric behaviour of the critical
point~\cite{n91b}.
\vskip.3cm

We made six Montecarlo simulations on lattices of sizes $(N_s)^2N_t$
for the values of
$\beta$ and $N_t$ described in tab.I . These values have been chosen in
order to check both the scaling as a function of $\beta$ and the
temperature dependence of the physical observables. In fact the first
three combinations: $(\beta=0.739,N_t=12)$,  $(\beta=0.742,N_t=13)$,
 $(\beta=0.746,N_t=15)$, roughly correspond to the same physical
temperature $T/\L\sim 0.92$ (see tab.I), while the second set:
$(\beta=0.739,N_t=13)$,  $(\beta=0.742,N_t=14)$,
 $(\beta=0.746,N_t=16)$ corresponds to $T/\L\sim 0.85$. These are the
lowest temperatures that we could reach with our precision.
\vskip.3cm

The consistency of our data with the zero temperature result tells us
that they are low enough to be unaffected by the
higher order corrections which appear near the deconfinement transition.
The size in the space directions was fixed to be $N_s=48$, and periodic
boundary conditions were chosen also in these directions. A standard
heat-bath algorithm was used to update links. We measured
 all the possible Polyakov line correlations, and in the two
experiments, $(\beta=0.739,N_t=13)$ and $(\beta=0.746,N_t=15)$,\footnote
{We have chosen the two extreme situations, both in $\beta$ and in $T$.}
we also measured Wilson loops (orthogonal to the time direction)
of sizes (R,L) in the range $2\leq R,L \leq 20$.
\vskip.3cm

The major problem we had to face was the critical slowing down. We have
already discussed this problem in ref.~\cite{n90}.
Comparing the present data with those of ref.~\cite{n90},
we can say that the problem is even more severe
with finite temperature simulations. Moreover, Polyakov line
correlations are in general more
self-correlated than Wilson loops. We kept under control this problem
separating the measure of two successive  Wilson loops and Polyakov
line correlations with 64 sweeps. We also developed a code so
as to scatter both in space
and (Montecarlo) time the measured correlations, trying to optimize not
only self-correlations but also cross-correlations. We think that this
procedure could allow a significative improvement in precision for any
LGT simulation. The typical run was composed by 200.000 sweeps of
thermalization and 960.000 sweeps during which 15.000 measures
of Polyakov line correlations and Wilson loops were taken.
\vskip .3 cm

The resulting
expectation values were fitted first with the pure area law and then
adding the  fermionic string quantum contribution
(eq.s(\ref{zs},\ref{zb})).  This correction being fixed, we do not add
new free parameters in the fitting
procedure, hence the two fits (with and without quantum
contribution) can be directly compared. In fitting the data we had
also to take into account the periodicity of the
lattice: the same correlation between two Polyakov lines could also be
obtained following topologically non-trivial paths, winding around the
compactified space directions. We made nonlinear fits summing over the
nearest winding numbers\footnote{The same problem can be ignored with
Wilson loops.}. So the fitting function was
\eq
\langle P(\vec x)P(\vec x+\vec R) \rangle = \sum_{i,j=-1}^{i,j=1}
{\rm e}^{-F(R_{i,j},N_t)}~~~,
\label{fit}
\en
with
\eq
R_{i,j}=\sqrt{(R_x+iN_s)^2+(R_y+jN_s)^2}~~~,
\en
$R_x$ and $R_y$ being the Euclidean components of the vector $\vec R$
and
\eq
F(R_{i,j},N_t)=\sigma~R_{i,j}N_t~+~cost~+F_{f,4}(R_{i,j},N_t)~~~.
\en
Being $N_t$  fixed, the constant term must be the same for all the
correlations.
Notice that this fit is definitely more constrained than the similar
one for  Wilson loops, since in this case we have only two free
parameters, namely $cost$ and $\sigma$, instead of the three of the
Wilson loop.

In fitting the data we
used the expansions of $F_{f,4}$ up to the sixth order\footnote{As a matter of
fact only the dominant term and the second order give significative
contributions for generic values of $z$, higher orders become relevant
only in the point $z=1$ where they allow the smooth connection of the
two asymptotic behaviours. Since this point turns out to be very
important in our fits, we decided to keep higher orders anyway.} in  $q$
(eq.s(\ref{zs},\ref{zb})) both in the region $z<1$ and in $z>1$.
The Wilson loops analysis was done as described in ref.~\cite{n90}.
\vskip.3cm

Before discussing our results let us make few general comments on the
data:
\begin{enumerate}
 \item
A first feature of the Polyakov line correlations is their impressively
good scaling behaviour
(which is shown in fig.2) even before any cutting or fitting
manipulation. The two sets of data which overlap in fig.2 are taken at
different values of $\beta$ and $N_t$ tuned so as to give the same
physical temperature. The tuning crucially involves the critical index
$\nu=0.63$ and tells us first, that we are in the scaling region and
second, that, as expected, all the data are controlled only by the two
physical scales $T$ and $\sigma$ .
 \item
As described in sect.3, we cannot expect to see any string effect in the
$R$ dependence of  Polyakov line correlations at large distances
($z>1$) (in this region the only signature of the string is in the
$T$ dependence of
$\sigma$ and it can be seen only by comparison with the Wilson loops,
being the two values of $T$ we explore are too near to give measurable
effects). On the contrary string effects should be visible for $z<1$
as deviations from the linear rising behaviour of the potential.
This is clearly visible in fig.3 . In this figure the potential $V(R)$,
defined as
\eq
V(R)=-\frac{1}{N_t}\log \langle P(x)P^\dagger(x+R) \rangle ~~~,
\label{pot}
\en
is plotted against the interquark distance $R$. The linear rising
behaviour is clearly visible for $z>1$
(in the example $z=1$ corresponds to $R=7.5$),
and similarly evident is the smooth onset  of  the $1/R$ correction for $z<
1$.
\end{enumerate}

\vskip 0.3cm
Let us now concentrate on the quantitative tests.
We made several fits, cutting off short range data, as described in
sect.2 and for the Wilson loops we cut off short range data
following the procedure
of ref.~\cite{n90}. The  result of these fits is a set of values of the
string tension $\sigma(\beta,N_t,z_{min})$ (in the case of Wilson loops
$z_{min}$ is substituted by $R_{min}$), which depend on $\beta$, on the
temperature $T=1/N_ta$ and on the  threshold $z_{min}$ ($R_{min}$).
Wilson loops data are listed in tab.II together with two sets of data
taken from ref.\cite{n90} for comparison. The Polyakov line correlation
data are listed in tab.III . The errors on  $\sigma$ have been estimated
with an ordinary jackknife procedure, and the data are presented taking
into account their scaling behaviour in $\beta$:
\eq
k=\frac{\sigma}{(\beta_c-\beta)^{2\nu}}~~.
\en

Let us examine first the data coming from Wilson loops. We can see from
tab.II that:
\begin{description}
\item
If we add the fermionic correction, all data agree among them:
for all the sets, acceptable confidence levels are obtained already from
$R_{min}=3$, and, starting from the same threshold, scaling is fulfilled.

\item
On the contrary, if we neglect quantum corrections, acceptable
confidence levels with the area law are reached only at $R_{min}=5$,
and, what is more important, {\it scaling is never reached}.

\item
It is quite interesting to see that the data show no dependence at all
from the finite temperature. This is particularly impressive for the
results obtained without the string corrections: the differences between
various $\beta$'s due to the scaling violations are much bigger than the
difference between zero temperature and finite low temperature data for
nearby values of $\beta$: the data at ($\beta=0.739,~T/\L=0.84$) are much
more similar to those at ($\beta=0.740,~T=0$) than to those at
($\beta=0.746,~T/\L=0.92$), and similarly for the other pair
($\beta=0.746,~T/\L=0.92$) and ($\beta=0.745,~T=0$).

\item
The   values of
$\sigma$ obtained with the string corrections still show a decreasing
trend in the region $R_{min}>3$. This behaviour was already
noticed and discussed in ref.~\cite{n90}: it can be ascribed to higher order
contributions of the string and can be successfully fitted with a root
law~\cite{n90}. The important point is that this trend (as a difference
with the no-string case)  becomes
asymptotic already in the region $R_{min}=5,6$ which we can reach with
our measures. Since the new data are in agreement with those at $T=0$,
we can borrow the result of ref.~\cite{n90} and give
\eq
k_w\equiv \frac{\sigma_{wilson}}{(\beta_c-\beta)^{2\nu}}=3.65(9)
\label{kw}
\en

as a final value for the string tension extracted from Wilson loops.
\end{description}

Analyzing now the data coming from Polyakov line correlations
(tab.III) we can see that:

\begin{description}
\item
In general, for the string-corrected fits, acceptable confidence levels
are  obtained for $z\geq z_{min}=0.6$ , while using the pure area low
one finds acceptable CL only for  $z\geq z_{min}=0.8$~.

\item
The fits to the scaling law show the same thresholds and give definitely
good confidence levels for $z\geq z_{min}=1.2$, a region in which
the string correction only acts as a renormalization of $\sigma$
(see eq.(\ref{zb})). However
the fact that the fermionic string  describes better the data can be
still stressed observing that, after $z=1.0$, the
string-corrected data give a stable $\sigma$, while those obtained with
the pure area law keep decreasing.

\item
Taking as final answer for the string tension the value (and error)
extracted at $z_{min}=1.2$, we have\footnote {The absolute value of
$\sigma$ does not change within the
errors taking any value of $z_{min}\geq 1.0$. Only the error is a function
of $z_{min}$ and we  choose , of course, the $z$ value corresponding to
the maximal confidence level, which is higher than 70\%.}
\eq
k_p\equiv \frac{\sigma_{polyakov}}{(\beta_c-\beta)^{2\nu}}=3.70(4)
\en
which is in impressive and remarkable agreement with the value $k_w$
given in eq.(\ref{kw}).

\end{description}

Let us conclude  making the same comparison between Wilson and
Po\-lya\-kov data, without taking into account string corrections.

Since in this case
Wilson loops data do not scale and Polyakov data should depend on $N_t$,
 we cannot do this comparison at the
level of scaling variables $k_w$ and $k_p$ as we did above, but we must
examine the data at each value of $\beta$ and $N_t$ separately.
The result is shown
in fig.4 and fig.5: there is clearly a gap between the values of
$\sigma$ evaluated with these two different observables.
\vskip .3 cm

In order to get a numerical estimate of this gap, we
assumed the following rule: consider, both for Wilson loops  and for
 Polyakov line correlations, the first value of the string tension
(as $z_{min}$ and $R_{min}$ increase)
with a  good confidence level of the fit to the pure area law.
As discussed above, this means that we must take $R_{min}=5$ and $z_{min}
=0.8$. With this criterion we obtain:

$\sigma(N_t=13,\beta=0.739~,~R_{min}=5)_{wilson}=0.0338(7)~~,$

$\sigma(N_t=13,\beta=0.739~,~z_{min}=0.8)_{polyakov}=0.0315(4)~~,$

with a gap 0.0023(11)~;

$\sigma(N_t=15,\beta=0.746~,~R_{min}=5)_{wilson}=0.0214(3)~~,$

$\sigma(N_t=15,\beta=0.746,z_{min}=1.2)_{polyakov}=0.0195(3)~~,$

with a gap 0.0019(6)~.

Remarkably, both are compatible with a value $\tilde c=1$ for the
effective conformal anomaly. Indeed, since these gaps should be due,
at least
asymptotically, to the term $\frac{\tilde c \pi}{6 N_t^2}$ of
eq.(\ref{zb}), setting
$\tilde c=1$, we expect a gap of 0.0031 for $N_t=13$ and of
0.0023 for $N_t=15$.

Needless to say that these numbers  must be taken with caution, since both
the $\sigma_p$ and $\sigma_w$ are still moving toward their asymptotic
values in the region that we are measuring. Notwithstanding this, they
clearly indicate the presence of a gap between the two estimates of
$\sigma$, and this fact can simply be understood only in the context
of an effective string picture of the infrared behaviour of the gauge
theories.


\vskip .3 cm
\hrule
\vskip .7 cm
\centerline{\sl Figure Captions}
\vskip .3 cm
\begin{description}

\item{Fig.1)}
The string tension $\sigma$ as a function of $z_{min}$ for the fermion,
 boson  and no-string pictures at $\beta=0.746, N_t=15$.
\item{Fig.2)}
Polyakov line correlations as functions of the adimensional ratio $z=2R
/N_t$. Three samples of data are presented: $\beta=0.746, N_t=16$,
$\beta=0.742, N_t=14$ and $\beta=0.742, N_t=13$. The first two, which
overlap in the figure, correspond to the same physical temperature $T$.
The third one, with error bars, is plotted for comparison.

\item{Fig.3)}
The interquark potential $V(R)$ defined in eq.(\ref{pot}) as a function
of the distance  $R$. The data are taken from the
sample at $\beta=0.746, N_t=15$, error bars are presented only for
$R>8$. For $R<8$ errors are smaller than the plotted symbols. The
straight line corresponds to the pure area law fit. The dashed line is
the fit with the fermionic string correction while the dotted line is
the fit with the bosonic string correction.

\item{Fig.4)}
String tension extracted from the Polyakov line correlations (square)
and from Wilson loops (diamonds) using a pure area law, without string
corrections, plotted  as
functions of the short range cutoff $R_{min}$ .
The sample of data is at $\beta=0.746, N_t=15$. The values of $z_{min}$
for the Polyakov case are converted using the relation $R=z\cdot N_t/2$.

\item{Fig.5)}
Same as Fig.4, taking into account the fermionic string corrections.

\end{description}

\newpage

\centerline{\sl Table Captions}
\vskip .3 cm
\begin{description}
\item{Tab.I}
Set of measured data: in the third column the corresponding physical
temperatures are presented. In the last column we list the samples in
which also Wilson loops were measured.

\item{Tab.II}
Values of string tensions extracted from Wilson loops as functions
of $R_m$
for the two set of data: $\beta=0.739$, $\beta=0.746$  presented in this
paper and for three sets of data of ref.~\cite{n90}: $\beta=0.740$,
$\beta=0.745$, $\beta=0.750$. The string tension is presented in scaled
unities in the case of the pure area law fits (IIa) and taking into
account  fermionic string corrections (IIb). In both tables the sixth
line contains the result (k) of the scaling fit and the last line the
corresponding reduced $\chi^2$ and confidence levels.

\item{Tab.III}
Values of string tensions extracted from Polyakov line correlations as
functions of $z_{min}$.
The string tension is presented in scaled
unities in the case of the pure area law fits (IIIa) and taking into
account  fermionic string corrections (IIIb).
In both tables the last two
lines contain the result (k) of the scaling fit and the
corresponding reduced $\chi^2$ and confidence levels.

\end{description}

\newpage

\scriptsize
\centerline {\bf Tab. I}
\vskip.5cm
$$\vbox{\offinterlineskip \halign
{&\vrule#&\hfil\quad # \quad\quad\hfil
&\vrule#&\hfil\quad # \quad\hfil
&\vrule#&\hfil\quad # \quad\hfil
&\vrule#&\hfil\quad # \quad\hfil
&\vrule#\cr
\noalign{\hrule}
height3pt&\hfil&&\hfil&&\hfil&&\hfil&\cr
&$\beta$&&$N_t$&&$T/\Lambda$&&$Wilson$&\cr
height3pt&\hfil&&\hfil&&\hfil&&\hfil&\cr
\noalign{\hrule}
height3pt&\hfil&&\hfil&&\hfil&&\hfil&\cr
&0.739&&12&&0.91&&\quad&\cr
height3pt&\hfil&&\hfil&&\hfil&&\hfil&\cr
\noalign{\hrule}
height3pt&\hfil&&\hfil&&\hfil&&\hfil&\cr
&0.739&&13&&0.84&&$\times$&\cr
height3pt&\hfil&&\hfil&&\hfil&&\hfil&\cr
\noalign{\hrule}
height3pt&\hfil&&\hfil&&\hfil&&\hfil&\cr
&0.742&&13&&0.92&&\quad&\cr
height3pt&\hfil&&\hfil&&\hfil&&\hfil&\cr
\noalign{\hrule}
height3pt&\hfil&&\hfil&&\hfil&&\hfil&\cr
&0.742&&14&&0.86&&\quad&\cr
height3pt&\hfil&&\hfil&&\hfil&&\hfil&\cr
\noalign{\hrule}
height3pt&\hfil&&\hfil&&\hfil&&\hfil&\cr
&0.746&&15&&0.92&&$\times$&\cr
height3pt&\hfil&&\hfil&&\hfil&&\hfil&\cr
\noalign{\hrule}
height3pt&\hfil&&\hfil&&\hfil&&\hfil&\cr
&0.746&&16&&0.87&&\quad&\cr
height3pt&\hfil&&\hfil&&\hfil&&\hfil&\cr
\noalign{\hrule}
}}$$
\vskip1.cm


\centerline {\bf Tab. II (a)}
\vskip.5cm
$$\vbox{\offinterlineskip \halign
{&\vrule#
 &\hfil # \hfil
 &\vrule#&\quad # \quad\hfil
 &\vrule#&\quad # \quad\hfil
 &\vrule#&\quad # \quad\hfil
 &\vrule#&\quad # \quad\hfil
 &\vrule#&\quad # \quad\hfil
 &\vrule#\cr
\noalign{\hrule}
\noalign{\hrule}
height3pt
&\hfil&
&\hfil&
&\hfil&
&\hfil&
&\hfil&
&\hfil&\cr
&$\beta$&
&$R_{min}=2$&
&\quad\quad$3$&
&\quad\quad$4$&
&\quad\quad$5$&
&\quad\quad$6$&\cr
height3pt
&\hfil&
&\hfil&
&\hfil&
&\hfil&
&\hfil&
&\hfil&\cr
\noalign{\hrule}
height4pt&\hfil&&\hfil&&\hfil&&\hfil&&\hfil&&\hfil&\cr
&0.739&&5.03(1)&&4.50(2)&&4.22(4)&&4.06(8)&&4.14(16)&\cr
height4pt&\hfil&&\hfil&&\hfil&&\hfil&&\hfil&&\hfil&\cr
\noalign{\hrule}
height4pt&\hfil&&\hfil&&\hfil&&\hfil&&\hfil&&\hfil&\cr
&0.740&&4.90(1)&&4.47(3)&&4.24(5)&&4.05(9)&&4.09(19)&\cr
height4pt&\hfil&&\hfil&&\hfil&&\hfil&&\hfil&&\hfil&\cr
\noalign{\hrule}
height4pt&\hfil&&\hfil&&\hfil&&\hfil&&\hfil&&\hfil&\cr
&0.745&&4.97(2)&&4.51(3)&&4.30(5)&&4.17(9)&&3.99(16)&\cr
height4pt&\hfil&&\hfil&&\hfil&&\hfil&&\hfil&&\hfil&\cr
\noalign{\hrule}
height4pt&\hfil&&\hfil&&\hfil&&\hfil&&\hfil&&\hfil&\cr
&0.746&&5.14(1)&&4.62(2)&&4.36(3)&&4.12(6)&&4.20(9)&\cr
height4pt&\hfil&&\hfil&&\hfil&&\hfil&&\hfil&&\hfil&\cr
\noalign{\hrule}
height4pt&\hfil&&\hfil&&\hfil&&\hfil&&\hfil&&\hfil&\cr
&0.750&&5.30(3)&&4.77(6)&&4.46(8)&&4.32(14)&&4.21(25)&\cr
height4pt&\hfil&&\hfil&&\hfil&&\hfil&&\hfil&&\hfil&\cr
\noalign{\hrule}
\noalign{\hrule}
height4pt&\hfil&&\hfil&&\hfil&&\hfil&&\hfil&&\hfil&\cr
&$k$&&5.028(5)&&4.55(1)&&4.31(2)&&4.12(4)&&4.14(6)&\cr
height4pt&\hfil&&\hfil&&\hfil&&\hfil&&\hfil&&\hfil&\cr
\noalign{\hrule}
height4pt&\hfil&&\hfil&&\hfil&&\hfil&&\hfil&&\hfil&\cr
&$\chi^2_{red}~;~C.L.$&& 35.0~;~0.0&&10.2~;~0.0&
& 3.3~;~0.0&&0.9~;~0.5&
& 0.4~;~0.8&\cr
height4pt&\hfil&&\hfil&&\hfil&&\hfil&&\hfil&&\hfil&\cr
\noalign{\hrule}
\noalign{\hrule}}}$$
\vskip1.cm


\centerline {\bf Tab. II (b)}
\vskip.5cm
$$\vbox{\offinterlineskip \halign
{&\vrule#
 &\hfil # \hfil
 &\vrule#&\quad # \quad\hfil
 &\vrule#&\quad # \quad\hfil
 &\vrule#&\quad # \quad\hfil
 &\vrule#&\quad # \quad\hfil
 &\vrule#&\quad # \quad\hfil
 &\vrule#\cr
\noalign{\hrule}
\noalign{\hrule}
height3pt
&\hfil&
&\hfil&
&\hfil&
&\hfil&
&\hfil&
&\hfil&\cr
&$\beta$&
&$R_{min}=2$&
&\quad\quad$3$&
&\quad\quad$4$&
&\quad\quad$5$&
&\quad\quad$6$&\cr
height3pt
&\hfil&
&\hfil&
&\hfil&
&\hfil&
&\hfil&
&\hfil&\cr
\noalign{\hrule}
height4pt&\hfil&&\hfil&&\hfil&&\hfil&&\hfil&&\hfil&\cr
&0.739&&4.26(1)&&4.01(2)&&3.87(4)&&3.79(8)&&3.92(16)&\cr
height4pt&\hfil&&\hfil&&\hfil&&\hfil&&\hfil&&\hfil&\cr
\noalign{\hrule}
height4pt&\hfil&&\hfil&&\hfil&&\hfil&&\hfil&&\hfil&\cr
&0.740&&4.19(1)&&4.01(3)&&3.90(5)&&3.77(9)&&3.72(19)&\cr
height4pt&\hfil&&\hfil&&\hfil&&\hfil&&\hfil&&\hfil&\cr
\noalign{\hrule}
height4pt&\hfil&&\hfil&&\hfil&&\hfil&&\hfil&&\hfil&\cr
&0.745&&4.15(2)&&3.98(3)&&3.91(5)&&3.85(9)&&3.73(16)&\cr
height4pt&\hfil&&\hfil&&\hfil&&\hfil&&\hfil&&\hfil&\cr
\noalign{\hrule}
height4pt&\hfil&&\hfil&&\hfil&&\hfil&&\hfil&&\hfil&\cr
&0.746&&4.23(1)&&4.04(2)&&3.93(3)&&3.77(6)&&3.89(9)&\cr
height4pt&\hfil&&\hfil&&\hfil&&\hfil&&\hfil&&\hfil&\cr
\noalign{\hrule}
height4pt&\hfil&&\hfil&&\hfil&&\hfil&&\hfil&&\hfil&\cr
&0.750&&4.24(3)&&4.07(6)&&3.96(8)&&3.90(14)&&3.84(25)&\cr
height4pt&\hfil&&\hfil&&\hfil&&\hfil&&\hfil&&\hfil&\cr
\noalign{\hrule}
\noalign{\hrule}
height4pt&\hfil&&\hfil&&\hfil&&\hfil&&\hfil&&\hfil&\cr
&$k$&&4.221(5)&&4.02(1)&&3.91(2)&&3.80(4)&&3.85(6)&\cr
height4pt&\hfil&&\hfil&&\hfil&&\hfil&&\hfil&&\hfil&\cr
\noalign{\hrule}
height4pt&\hfil&&\hfil&&\hfil&&\hfil&&\hfil&&\hfil&\cr
&$\chi^2_{red}~;~C.L.$&& 9.7~;~0.0&&0.9~;~0.4&
& 0.5~;~0.7&&0.3~;~0.9&
& 0.3~;~0.8&\cr
height4pt&\hfil&&\hfil&&\hfil&&\hfil&&\hfil&&\hfil&\cr
\noalign{\hrule}
\noalign{\hrule}}}$$

\newpage


\centerline{\bf Tab. III (a)}
\vskip 0.5cm
\vbox{\offinterlineskip \halign
{&\vrule#&#&\vrule#&#&\vrule#&#&\vrule#&#&
\vrule#&#&\vrule#&#&\vrule#&#&\vrule#&#&\vrule#&#&\vrule#\cr
\noalign{\hrule}
height5pt&\multispan3 \hfil&&\multispan{13} \hfil&\cr
&\multispan2 \hfil&\hfil  $z_{min}$ \quad \hfil&&
\hfil\quad 0.2 \quad\hfil&\omit&\hfil\quad 0.4 \quad\hfil&\omit&
\hfil\quad 0.6 \quad\hfil&\omit&\hfil\quad 0.8 \quad\hfil&\omit&
\hfil\quad 1.0 \quad\hfil&\omit&\hfil\quad 1.2 \quad\hfil&\omit&
\hfil\quad 1.4 \quad\hfil&\cr
height2pt&\multispan3 \hfil&&\multispan{13} \hfil&\cr
&\hfil\quad $\beta$ \quad\hfil&\omit
&\hfil\quad $N_t$ \quad\hfil
&\hfil&&\multispan{12}\hfil&\cr
height2pt&\multispan3 \hfil&&\multispan{13} \hfil&\cr
\noalign{\hrule}
height5pt&\multispan3 \hfil&&\hfil&&\hfil&&
\hfil&&\hfil&&\hfil&&\hfil&&\hfil&\cr
&\hfil\quad 0.739 \quad\hfil&\omit&~~12\hfil&&
\hfil\quad 4.48(3) \quad\hfil&&\hfil\quad 4.04(3) \quad\hfil&&
\hfil\quad 3.89(4) \quad\hfil&&\hfil\quad 3.79(5) \quad\hfil&&
\hfil\quad 3.71(5) \quad\hfil&&\hfil\quad 3.64(7) \quad\hfil&&
\hfil\quad 3.64(12) \quad\hfil&\cr
height3pt&\multispan3 \hfil&&\hfil&&\hfil&&
\hfil&&\hfil&&\hfil&&\hfil&&\hfil&\cr
&\multispan3 \hfil&&\multispan{13} \hrulefill&\cr
height3pt&\multispan3 \hfil&&\hfil&&\hfil&&
\hfil&&\hfil&&\hfil&&\hfil&&\hfil&\cr
&\hfil\quad 0.739 \quad\hfil&\omit&~~13\hfil&&
\hfil\quad 4.63(2) \quad\hfil&&\hfil\quad 4.15(3) \quad\hfil&&
\hfil\quad 3.93(4) \quad\hfil&&\hfil\quad 3.78(4) \quad\hfil&&
\hfil\quad 3.67(6) \quad\hfil&&\hfil\quad 3.57(12) \quad\hfil&&
\hfil\quad 3.54(24) \quad\hfil&\cr
height3pt&\multispan3 \hfil&&\hfil&&\hfil&&
\hfil&&\hfil&&\hfil&&\hfil&&\hfil&\cr
&\multispan3 \hfil&&\multispan{13} \hrulefill&\cr
height3pt&\multispan3 \hfil&&\hfil&&\hfil&&
\hfil&&\hfil&&\hfil&&\hfil&&\hfil&\cr
&\hfil\quad 0.742 \quad\hfil&\omit&~~13\hfil&&
\hfil\quad 4.58(2) \quad\hfil&&\hfil\quad 4.12(3) \quad\hfil&&
\hfil\quad 3.94(4) \quad\hfil&&\hfil\quad 3.83(5) \quad\hfil&&
\hfil\quad 3.76(7) \quad\hfil&&\hfil\quad 3.72(10) \quad\hfil&&
\hfil\quad 3.69(17) \quad\hfil&\cr
height3pt&\multispan3 \hfil&&\hfil&&\hfil&&
\hfil&&\hfil&&\hfil&&\hfil&&\hfil&\cr
&\multispan3 \hfil&&\multispan{13} \hrulefill&\cr
height3pt&\multispan3 \hfil&&\hfil&&\hfil&&
\hfil&&\hfil&&\hfil&&\hfil&&\hfil&\cr
&\hfil\quad 0.742 \quad\hfil&\omit&~~14\hfil&&
\hfil\quad 4.65(1) \quad\hfil&&\hfil\quad 4.16(2)  \quad\hfil&&
\hfil\quad 3.91(3) \quad\hfil&&\hfil\quad 3.76(3) \quad\hfil&&
\hfil\quad 3.64(3) \quad\hfil&&\hfil\quad 3.57(9) \quad\hfil&&
\hfil\quad 3.56(14) \quad\hfil&\cr
height3pt&\multispan3 \hfil&&\hfil&&\hfil&&
\hfil&&\hfil&&\hfil&&\hfil&&\hfil&\cr
&\multispan3 \hfil&&\multispan{13} \hrulefill&\cr
height3pt&\multispan3 \hfil&&\hfil&&\hfil&&
\hfil&&\hfil&&\hfil&&\hfil&&\hfil&\cr
&\hfil\quad 0.746 \quad\hfil&\omit&~~15\hfil&&
\hfil\quad 4.36(3) \quad\hfil&&\hfil\quad 4.08(3) \quad\hfil&&
\hfil\quad 3.83(5) \quad\hfil&&\hfil\quad 3.75(6) \quad\hfil&&
\hfil\quad 3.65(8) \quad\hfil&&\hfil\quad 3.62(10) \quad\hfil&&
\hfil\quad 3.56(13) \quad\hfil&\cr
height3pt&\multispan3 \hfil&&\hfil&&\hfil&&
\hfil&&\hfil&&\hfil&&\hfil&&\hfil&\cr
&\multispan3 \hfil&&\multispan{13} \hrulefill&\cr
height3pt&\multispan3 \hfil&&\hfil&&\hfil&&
\hfil&&\hfil&&\hfil&&\hfil&&\hfil&\cr
&\hfil\quad 0.746 \quad\hfil&\omit&~~16\hfil&&
\hfil\quad 4.52(3) \quad\hfil&&\hfil\quad 4.08(4) \quad\hfil&&
\hfil\quad 3.94(5) \quad\hfil&&\hfil\quad 3.82(6) \quad\hfil&&
\hfil\quad 3.81(9) \quad\hfil&&\hfil\quad 3.71(13) \quad\hfil&&
\hfil\quad 3.79(21) \quad\hfil&\cr
height3pt&\multispan3 \hfil&&\hfil&&\hfil&&
\hfil&&\hfil&&\hfil&&\hfil&&\hfil&\cr
&\multispan3 \hfil&&\multispan{13} \hrulefill&\cr
height3pt&\multispan3 \hfil&&\hfil&&\hfil&&
\hfil&&\hfil&&\hfil&&\hfil&&\hfil&\cr
&\hfil\quad  \quad\hfil&\omit $k$&\hfil&&
\hfil\quad 4.58(1) \quad\hfil&&\hfil\quad 4.12(1) \quad\hfil&&
\hfil\quad 3.91(2) \quad\hfil&&\hfil\quad 3.78(2) \quad\hfil&&
\hfil\quad 3.69(3) \quad\hfil&&\hfil\quad 3.63(4) \quad\hfil&&
\hfil\quad 3.62(6) \quad\hfil&\cr
height3pt&\multispan3 \hfil&&\hfil&&\hfil&&
\hfil&&\hfil&&\hfil&&\hfil&&\hfil&\cr
&\multispan3 \hfil&&\multispan{13} \hrulefill&\cr
height3pt&\multispan3 \hfil&&\hfil&&\hfil&&
\hfil&&\hfil&&\hfil&&\hfil&&\hfil&\cr
&\hfil\quad$\chi^2_{red}$  \quad\hfil&\omit&$;~~C.L.$\hfil&&
\hfil\quad 16.0~;~0.0 \quad\hfil&&\hfil\quad 3.5~;~0.0 \quad\hfil&&
\hfil\quad 0.7~;~0.5 \quad\hfil&&\hfil\quad 0.3~;~0.9 \quad\hfil&&
\hfil\quad 0.7~;~0.5 \quad\hfil&&\hfil\quad 0.3~;~0.9 \quad\hfil&&
\hfil\quad 0.2~;~0.9 \quad\hfil&\cr
height3pt&\multispan3 \hfil&&\hfil&&\hfil&&
\hfil&&\hfil&&\hfil&&\hfil&&\hfil&\cr
&\multispan3 \hfil&&\multispan{13} \hrulefill&\cr
\noalign{\hrule}}}
\vskip 1.cm


\centerline{\bf Tab. III (b)}
\vskip 0.5cm
\vbox{\offinterlineskip \halign
{&\vrule#&#&\vrule#&#&\vrule#&#&\vrule#&#&
\vrule#&#&\vrule#&#&\vrule#&#&\vrule#&#&\vrule#&#&\vrule#\cr
\noalign{\hrule}
height5pt&\multispan3 \hfil&&\multispan{13} \hfil&\cr
&\multispan2 \hfil&\hfil  $z_{min}$ \quad \hfil&&
\hfil\quad 0.2 \quad\hfil&\omit&\hfil\quad 0.4 \quad\hfil&\omit&
\hfil\quad 0.6 \quad\hfil&\omit&\hfil\quad 0.8 \quad\hfil&\omit&
\hfil\quad 1.0 \quad\hfil&\omit&\hfil\quad 1.2 \quad\hfil&\omit&
\hfil\quad 1.4 \quad\hfil&\cr
height2pt&\multispan3 \hfil&&\multispan{13} \hfil&\cr
&\hfil\quad $\beta$ \quad\hfil&\omit
&\hfil\quad $N_t$ \quad\hfil
&\hfil&&\multispan{12}\hfil&\cr
height2pt&\multispan3 \hfil&&\multispan{13} \hfil&\cr
\noalign{\hrule}
height5pt&\multispan3 \hfil&&\hfil&&\hfil&&
\hfil&&\hfil&&\hfil&&\hfil&&\hfil&\cr
&\hfil\quad 0.739 \quad\hfil&\omit&~~12\hfil&&
\hfil\quad 4.07(2) \quad\hfil&&\hfil\quad 3.89(3) \quad\hfil&&
\hfil\quad 3.80(4) \quad\hfil&&\hfil\quad 3.75(5) \quad\hfil&&
\hfil\quad 3.71(5) \quad\hfil&&\hfil\quad 3.67(7) \quad\hfil&&
\hfil\quad 3.70(12) \quad\hfil&\cr
height3pt&\multispan3 \hfil&&\hfil&&\hfil&&
\hfil&&\hfil&&\hfil&&\hfil&&\hfil&\cr
&\multispan3 \hfil&&\multispan{13} \hrulefill&\cr
height3pt&\multispan3 \hfil&&\hfil&&\hfil&&
\hfil&&\hfil&&\hfil&&\hfil&&\hfil&\cr
&\hfil\quad 0.739 \quad\hfil&\omit&~~13\hfil&&
\hfil\quad 4.17(2) \quad\hfil&&\hfil\quad 3.98(3) \quad\hfil&&
\hfil\quad 3.84(4) \quad\hfil&&\hfil\quad 3.74(4) \quad\hfil&&
\hfil\quad 3.67(6) \quad\hfil&&\hfil\quad 3.59(12) \quad\hfil&&
\hfil\quad 3.59(24) \quad\hfil&\cr
height3pt&\multispan3 \hfil&&\hfil&&\hfil&&
\hfil&&\hfil&&\hfil&&\hfil&&\hfil&\cr
&\multispan3 \hfil&&\multispan{13} \hrulefill&\cr
height3pt&\multispan3 \hfil&&\hfil&&\hfil&&
\hfil&&\hfil&&\hfil&&\hfil&&\hfil&\cr
&\hfil\quad 0.742 \quad\hfil&\omit&~~13\hfil&&
\hfil\quad 4.14(2) \quad\hfil&&\hfil\quad 3.96(3) \quad\hfil&&
\hfil\quad 3.86(3) \quad\hfil&&\hfil\quad 3.81(5) \quad\hfil&&
\hfil\quad 3.78(7) \quad\hfil&&\hfil\quad 3.77(10) \quad\hfil&&
\hfil\quad 3.77(17) \quad\hfil&\cr
height3pt&\multispan3 \hfil&&\hfil&&\hfil&&
\hfil&&\hfil&&\hfil&&\hfil&&\hfil&\cr
&\multispan3 \hfil&&\multispan{13} \hrulefill&\cr
height3pt&\multispan3 \hfil&&\hfil&&\hfil&&
\hfil&&\hfil&&\hfil&&\hfil&&\hfil&\cr
&\hfil\quad 0.742 \quad\hfil&\omit&~~14\hfil&&
\hfil\quad 4.16(1) \quad\hfil&&\hfil\quad 3.97(2)  \quad\hfil&&
\hfil\quad 3.82(3) \quad\hfil&&\hfil\quad 3.72(3) \quad\hfil&&
\hfil\quad 3.63(5) \quad\hfil&&\hfil\quad 3.59(9) \quad\hfil&&
\hfil\quad 3.61(14) \quad\hfil&\cr
height3pt&\multispan3 \hfil&&\hfil&&\hfil&&
\hfil&&\hfil&&\hfil&&\hfil&&\hfil&\cr
&\multispan3 \hfil&&\multispan{13} \hrulefill&\cr
height3pt&\multispan3 \hfil&&\hfil&&\hfil&&
\hfil&&\hfil&&\hfil&&\hfil&&\hfil&\cr
&\hfil\quad 0.746 \quad\hfil&\omit&~~15\hfil&&
\hfil\quad 4.04(3) \quad\hfil&&\hfil\quad 3.90(3) \quad\hfil&&
\hfil\quad 3.77(4) \quad\hfil&&\hfil\quad 3.72(6) \quad\hfil&&
\hfil\quad 3.67(8) \quad\hfil&&\hfil\quad 3.67(9) \quad\hfil&&
\hfil\quad 3.64(13) \quad\hfil&\cr
height3pt&\multispan3 \hfil&&\hfil&&\hfil&&
\hfil&&\hfil&&\hfil&&\hfil&&\hfil&\cr
&\multispan3 \hfil&&\multispan{13} \hrulefill&\cr
height3pt&\multispan3 \hfil&&\hfil&&\hfil&&
\hfil&&\hfil&&\hfil&&\hfil&&\hfil&\cr
&\hfil\quad 0.746 \quad\hfil&\omit&~~16\hfil&&
\hfil\quad 4.15(3) \quad\hfil&&\hfil\quad 3.93(4) \quad\hfil&&
\hfil\quad 3.86(5) \quad\hfil&&\hfil\quad 3.79(6) \quad\hfil&&
\hfil\quad 3.80(9) \quad\hfil&&\hfil\quad 3.74(13) \quad\hfil&&
\hfil\quad 3.84(21) \quad\hfil&\cr
height3pt&\multispan3 \hfil&&\hfil&&\hfil&&
\hfil&&\hfil&&\hfil&&\hfil&&\hfil&\cr
&\multispan3 \hfil&&\multispan{13} \hrulefill&\cr
height3pt&\multispan3 \hfil&&\hfil&&\hfil&&
\hfil&&\hfil&&\hfil&&\hfil&&\hfil&\cr
&\hfil\quad  \quad\hfil&\omit $k$&\hfil&&
\hfil\quad 4.14(1) \quad\hfil&&\hfil\quad 3.95(1) \quad\hfil&&
\hfil\quad 3.82(2) \quad\hfil&&\hfil\quad 3.75(2) \quad\hfil&&
\hfil\quad 3.69(3) \quad\hfil&&\hfil\quad 3.67(4) \quad\hfil&&
\hfil\quad 3.68(6) \quad\hfil&\cr
height3pt&\multispan3 \hfil&&\hfil&&\hfil&&
\hfil&&\hfil&&\hfil&&\hfil&&\hfil&\cr
&\multispan3 \hfil&&\multispan{13} \hrulefill&\cr
height3pt&\multispan3 \hfil&&\hfil&&\hfil&&
\hfil&&\hfil&&\hfil&&\hfil&&\hfil&\cr
&\hfil\quad$\chi^2_{red}$  \quad\hfil&\omit&$;~~C.L.$\hfil&&
\hfil\quad 4.0~;~0.0 \quad\hfil&&\hfil\quad 1.3~;~0.2 \quad\hfil&&
\hfil\quad 0.6~;~0.6 \quad\hfil&&\hfil\quad 0.4~;~0.8 \quad\hfil&&
\hfil\quad 0.8~;~0.5 \quad\hfil&&\hfil\quad 0.4~;~0.8 \quad\hfil&&
\hfil\quad 0.2~;~0.9 \quad\hfil&\cr
height3pt&\multispan3 \hfil&&\hfil&&\hfil&&
\hfil&&\hfil&&\hfil&&\hfil&&\hfil&\cr
&\multispan3 \hfil&&\multispan{13} \hrulefill&\cr
\noalign{\hrule}}}
\end{document}